\documentclass[12pt, a4paper]{article}
\pdfoutput=1
\usepackage[dvipdfmx]{graphicx}
\usepackage{amssymb}
\usepackage{amsmath}
\usepackage{bm}
\usepackage[table,xcdraw,dvipsnames]{xcolor} %for \rowcolors
\usepackage{cite}
\usepackage{slashed}
\usepackage{subfigure}
\usepackage{epstopdf}            
\usepackage{epsfig}
\usepackage{here}
\usepackage{comment}
\usepackage{booktabs} %for \toprule
\usepackage{colortbl} %for \rowcolor
\usepackage{mathptmx} % change all font
\DeclareMathAlphabet{\mathcal}{OMS}{cmsy}{m}{n} % change \mathcal 

\graphicspath{{figs/}} % Automatically look in the figs directory
\renewcommand{\thefootnote}{\fnsymbol{footnote}}
\numberwithin{equation}{section} % Eq.(Sec.eq.)
\def\beq#1\eeq{\begin{align}#1\end{align}}
\newcommand{\la}{\lambda}

\newcommand{\BR}[1]{\mathrm{BR}(#1)}

\newcommand{\SM}{{\rm SM}}
\newcommand{\DM}{{\rm DM}}
\newcommand{\GeV}{{\rm GeV}}
\newcommand{\MeV}{{\rm MeV}}
\newcommand{\rel}{{\rm rel}}
\newcommand{\eff}{{\rm eff}}
\newcommand{\self}{{\rm self}}
\newcommand{\hc}{\text{h.c.}}
\newcommand{\Hch}{{H^{\pm}}}

\usepackage{caption}
\definecolor{BlueViolet}{rgb}{0.2, 0.00, 0.7}
\definecolor{Blue}{rgb}{0.15, 0.00, 0.9}
\definecolor{light_blue}{rgb}{0.15, 0.35, 0.95}
\definecolor{kit_green}{rgb}{0, 
0.58823 %150/255
, 0.50980 %130/255
}
\usepackage[%dvipdfmx,
colorlinks=true, linkcolor=light_blue,citecolor=light_blue,urlcolor=kit_green]{hyperref} 

\begin{document}
\sloppy %https://tex.stackexchange.com/questions/9107/how-can-i-make-my-text-never-go-over-the-right-margin-by-always-hyphenating-or-b

\begin{titlepage}
\begin{center}

\hfill{P3H--23--073, TTP23--047}

\vskip 0.5in

{\LARGE{\bf 
Light mass window of inert doublet dark matter with lepton portal interaction
}}\\

\vskip .5in

{\large{Ryo Higuchi$^1$, Syuhei Iguro$^{2,3}$, Shohei Okawa$^4$, and Yuji Omura$^1$}}

\vskip 0.5cm

{\it $^1$
Department of Physics, Kindai University, Higashi-Osaka, Osaka 577-8502, Japan}\\[3pt]

{\it $^2$ Institute for Theoretical Particle Physics (TTP), Karlsruhe Institute of Technology (KIT),
Engesserstra{\ss}e 7, 76131 Karlsruhe, Germany}\\[3pt]
{\it $^3$ Institute for Astroparticle Physics (IAP),
Karlsruhe Institute of Technology (KIT), 
Hermann-von-Helmholtz-Platz 1, 76344 Eggenstein-Leopoldshafen, Germany}\\[3pt]

{\it $^4$ 
Departament de F\'isica Qu\`antica i Astrof\'isica, Institut de Ci\`encies del Cosmos (ICCUB),
Universitat de Barcelona, Mart\'i i Franqu\`es 1, E-08028 Barcelona, Spain
}\\[3pt]

\end{center}
\vskip .2in

%%%%%%%%%%%%%%%%%%%%%%%%%
\begin{abstract}
We study phenomenology of a light scalar dark matter (DM).
In the model, there are an inert doublet scalar and a singlet Dirac fermion $\psi$, both charged under a global $Z_2$ symmetry. 
The mass of the lightest inert scalar $H$ can be lighter than 10\,GeV by imposing appropriate relations between three scalar quartic couplings. 
The lightest $Z_2$ odd particle is stable and DM. 
In this paper, focusing on the parameter space where $H$ is lighter than $\psi$ and is DM, 
we discuss DM physics related to relic density, direct detection, indirect detection, collider searches and other cosmological observations.
We clarify differences from the case where $\psi$ is instead DM, which has been focused on in the previous works. 
\end{abstract}

%%%%%%%%%%%%%%%%%%%%%%%%%
\end{titlepage}

\setcounter{page}{1}
\renewcommand{\thefootnote}{\#\arabic{footnote}}
\setcounter{footnote}{0}

%%%%%%%%%%%%%%%%%%%%%%%%%
% Contents
%%%%%%%%%%%%%%%%%%%%%%%%%
\hrule
\tableofcontents
\vskip .2in
\hrule
\vskip .4in
%%%%%%%%%%%%%%%%%%%%%%%%%

%%%%%%%%%%%%%%%%%%%%%%%%%%%%%%%
\section{Introduction}
\label{sec:intro}
%%%%%%%%%%%%%%%%%%%%%%%%%%%%%%%
Dark matter (DM) with mass below the GeV scale has drawn significant attention in recent years, given a continuous progress in DM direct detection experiments using nuclear recoils. 
The experimental results put very strong limits on the traditional Weakly Interacting Massive Particle (WIMP) DM above 10\,GeV, while their detection techniques lose the sensitivity to lighter DM mass range.
To address DM detection in this light mass region, low-mass DM searches have been designed making use of electron recoils \cite{Essig:2012yx, Essig:2017kqs, LUX:2018akb, XENON:2019gfn, XENON:2019zpr, PandaX-II:2021nsg, LZ:2023poo}, 
cryogenic detectors \cite{SuperCDMS:2014cds, SuperCDMS:2015eex, SuperCDMS:2017nns, SuperCDMS:2018mne, SuperCDMS:2020aus, CRESST:1999ynq, CRESST-II:2014ezs, CRESST:2015txj, CRESST:2019jnq, CRESST:2022lqw, EDELWEISS:2020fxc, EDELWEISS:2022ktt}, 
CCDs \cite{Crisler:2018gci, SENSEI:2019ibb, SENSEI:2020dpa, DAMIC:2011khz, DAMIC:2016lrs, DAMIC:2019dcn, DAMIC:2020cut} and 
$p$-type point contact semiconductors \cite{CDEX:2018lau, CDEX:2020tkb}, 
which operate with much lower detection thresholds (see also Ref.~\cite{Essig:2015cda} for theoretical works). 
Other recent attempts to detect an irreducible component of light DM boosted due to cosmic-ray up-scatterings or solar reflection also provide new insights into the DM interactions with nucleons \cite{Bringmann:2018cvk, Cappiello:2019qsw, Bondarenko:2019vrb, Guo:2020drq, Ema:2020ulo, Alvey:2022pad, Kouvaris:2015nsa}, 
electrons \cite{Ema:2018bih, Dent:2020syp, Xia:2022tid, An:2017ojc} and  neutrinos \cite{Jho:2021rmn}.

Sub-GeV DM, if thermally produced, calls for light new mediator particles with mass well below the electroweak (EW) scale \cite{Boehm:2003hm, Fayet:2004bw}.
In a class of DM models, such light mediators have no standard model (SM) gauge interactions, arising from gauge singlet extensions of the SM.
Well-studied singlet mediators include 
dark photon \cite{Galison:1983pa, Holdom:1985ag, Foot:1991kb, Pospelov:2007mp, Arkani-Hamed:2008hhe}, 
dark Higgs \cite{Schabinger:2005ei, Patt:2006fw, Wells:2008xg, Batell:2009yf, Weihs:2011wp, Kim:2008pp, Pospelov:2011yp, Arkani-Hamed:2008hhe}, 
axion-like particles \cite{Nomura:2008ru, Dolan:2014ska, Gola:2021abm, Fitzpatrick:2023xks, Dror:2023fyd, Darme:2020sjf, Kamada:2017tsq, Hochberg:2018rjs, Bharucha:2022lty, Ghosh:2023tyz} and 
heavy neutral leptons \cite{Pospelov:2007mp, Batell:2017cmf, McKeen:2018pbb, Blennow:2019fhy, Biswas:2021kio, Li:2022bpp}.
In other DM models, light mediators can originate in EW multiplets. 
For example, a light neutral scalar $H$ comes from a scalar doublet field and plays a role of $t$-channel~\cite{Okawa:2020jea}\footnote{In Ref.~\cite{Boehm:2013jpa}, a similar $t$-channel light scalar mediator has been also discussed, based on a supersymmetric model.}
and $s$-channel mediators~\cite{Herms:2022nhd} in DM annihilation.
In Ref.~\cite{Iguro:2022tmr}, the discovery potential of such a light scalar in high-energy collider experiments is studied.
In Ref.~\cite{Herms:2023cyy}, it is shown that a particular linear combination of neutral scalars in two extra scalar doublets can have sufficiently light mass without any conflict with EW precision measurements and collier search bounds.
In these models, a gauge singlet field is added as a potential DM candidate, besides the extra scalar doublets. 
In Refs.~\cite{Okawa:2020jea, Iguro:2022tmr}, for instance, 
a Dirac fermion $\psi$ is introduced as a DM candidate and the phenomenology is studied, assuming that $\psi$ is the lightest in the models.

In this paper, we investigate another parameter space, where $H$ is lighter than $\psi$ and is DM, which is not studied in the previous works \cite{Okawa:2020jea, Iguro:2022tmr}. 
In Sec.~\ref{sec:model}, we summarize the model setups proposed in Refs.~\cite{Okawa:2020jea, Iguro:2022tmr} and show that the scalar quartic couplings are predicted to be ${\cal O}(1)$ and the charged scalar mass range is limited from a perturbativity discussion. 
A simple extension of that minimal realization to a less constrained setup is also introduced. 
In Sec.~\ref{sec:DM}, assuming $H$ is DM, we discuss the DM physics concerned with the thermal production, direct detection, indirect detection, self-scattering and a constraint from the effective number of neutrinos in cosmology. 
After some comments on other constraints from high-energy collider experiments and supernova 1987A in Sec.~\ref{sec:others}, we show in Sec.~\ref{sec:result} viable parameter space in the models as well as make a comparison with the case of $\psi$ being DM.
Sec.~\ref{Sec:summary} is devoted to summary.

%%%%%%%%%%%%%%%%%%%%%%%%%%%%%%%%%%%%%%%%%%%%%%%%%%%%%%%%%%%%
\section{Models}
\label{sec:model}
%%%%%%%%%%%%%%%%%%%%%%%%%%%%%%%%%%%%%%%%%%%%%%%%%%%%%%%%%%%%
In this section, we recapitulate the basic idea of realizing a light neutral scalar from the inert doublet in a minimal way, following \cite{Okawa:2020jea}.
We also show a simple extension that resolves a perturbativity problem in the minimal model. 

%%%%%%%%%%%%%%%%%%%%%%%%%%%%%%%%%%%%%%%%%%%%%%%%%%%%%%%%%%%
\begin{table}[t]
\begin{center}
\begin{tabular}{ccccccc}
\hline
\hline
Fields & spin   & ~~$SU(3)$~~ & ~~$SU(2)_L$~~  & ~~$U(1)_Y$~~   &   ~~$U(1)_L$~~   &   ~~$Z_2$~~    \\ \hline   
     $\ell^i_L$ & $1/2$ &  ${\bf 1}$        &${\bf 2}$ &   $-\frac{1}{2}$  &   1  &   $+$    \\ 
      $e^i_R$ & $1/2$ &  ${\bf 1}$        &${\bf 1}$ &   $-1$  &   1  &   $+$       \\   \hline
     $\psi_L$ & $1/2$ &  ${\bf 1}$        &${\bf 1}$ &   $0$  &   1  &   $-$     \\  
     $\psi_R$ & $1/2$ &  ${\bf 1}$        &${\bf 1}$ &   $0$  &   1  &   $-$     \\  \hline
     $\Phi$ & $1$ & ${\bf 1}$         &${\bf2}$ &   $\frac{1}{2}$  &   0  &   $+$      \\  
          $\Phi_\nu$ & $1$ & ${\bf 1}$         &${\bf2}$ &   $\frac{1}{2}$  &   0  &   $-$        \\ 
                       \hline \hline
\end{tabular}
\end{center}
\caption{Charge assignment of the relevant particles. $i=1, \, 2, \, 3$ denotes the lepton flavor. $U(1)_L$ and $Z_2$ are global symmetry.}
 \label{tab:matter}
\end{table} 
%%%%%%%%%%%%%%%%%%%%%%%%%%%%%%%%%%%%%%%%%%%

%%%%%%%%%%%%%%%%%%%%%%%%%%%%%%%
\subsection{Minimal model}
\label{sec:minimal}
%%%%%%%%%%%%%%%%%%%%%%%%%%%%%%%
The model features a SM gauge singlet Dirac fermion $\psi$ and an extra scalar doublet $\Phi_\nu$ that has the same gauge quantum number as the SM Higgs doublet field $\Phi$. 
The two scalar doublets are distinguished by an additional global $Z_2$ symmetry which also guarantees the stability of DM. The global $U(1)_L$ symmetry is also assigned to avoid radiative generation of the active neutrino masses. The charge assignment is summarized in Table \ref{tab:matter}.

%%%%%%%%%%%%%%%%%%%%%%%%%%%%%%%
\subsubsection{Scalar sector}
\label{sec:scalar}
%%%%%%%%%%%%%%%%%%%%%%%%%%%%%%%
The scalar potential of the model is given by 
\begin{align}
V & = m_{11}^2 (\Phi^\dagger \Phi) 
   + m_{22}^2 (\Phi_\nu^\dagger \Phi_\nu) 
   + \lambda_1 (\Phi^\dagger \Phi)^2 
   + \lambda_2 (\Phi_\nu^\dagger \Phi_\nu)^2 \nonumber  \\
   & \quad + \lambda_3 (\Phi^\dagger \Phi) (\Phi_\nu^\dagger \Phi_\nu)
   + \lambda_4 (\Phi^\dagger \Phi_\nu) (\Phi_\nu^\dagger \Phi) 
   + \frac{1}{2} \lambda_5 [ (\Phi^\dagger \Phi_\nu)^2 + \hc ] ,
\label{eq:potential}
\end{align}
where all parameters are chosen to be real using the phase redefinition of $\Phi$ and $\Phi_\nu$.
We assume in this paper that only $\Phi$ develops a non-zero vacuum expectation value (VEV) and $\Phi_\nu$ does not. 
The scalar doublets are then expressed in terms of physical degrees of freedom and Nambu-Goldstone bosons $G^{0,\pm}$ as 
\begin{align}
\Phi = \begin{pmatrix} G^+ \\ \frac{1}{\sqrt{2}} (v+h+iG^0) \end{pmatrix} ,\quad 
\Phi_\nu = \begin{pmatrix} H^+ \\ \frac{1}{\sqrt{2}} (H+iA) \end{pmatrix} ,
\end{align}
where $v \simeq 246$\,GeV. 
After the EW symmetry breaking, the scalar masses are expressed as
\begin{align}
m_h^2 & = 2 \lambda_1 v^2 , \label{eq;Mh} \\
m_{H^\pm}^2 & = m_{22}^2 + \frac{\lambda_3 v^2}{2} , \label{eq:MHC} \\
m_A^2 & = m_\Hch^2 + \frac{(\lambda_4-\lambda_5) v^2}{2} , \label{eq:MA} \\
m_H^2 & = m_\Hch^2 +\frac{(\lambda_4+\lambda_5) v^2}{2} . \label{eq:MH}
\end{align}
We see that the mass splittings between the extra scalars are generated only by the quartic couplings.

Hereafter we consider the case where $m_H \leq m_A$\footnote{This assumption does not lose the generality of discussion since this is just a difference of the $\Phi_\nu$ basis. Rephasing $\Phi_\nu \to i \Phi_\nu$ interchanges the role of $H$ and $A$, which does not generate any physical difference.}
and derive three basic conditions for $H$ to be lighter than 10\,GeV.
The LEP result restricts the charged scalar mass to be $m_{H^\pm} > 70\text{--}90\,\GeV$~\cite{Pierce:2007ut} and excludes the neutral scalar mass region satisfying all three inequalities~\cite{Lundstrom:2008ai}: 
\begin{equation}
    m_H < 80\,\GeV, \quad m_A < 100\,\GeV, \quad m_A - m_H > 8\,\GeV.
\end{equation}
In addition, the EW precision data, especially the oblique $T$ parameter, restricts the mass difference between $A$ and $H^\pm$. The upper bound is estimated as $|m_{H^\pm}-m_A| \lesssim {\cal O}(10)\,\GeV$, which implies $\la_4\simeq\la_5$.
Thus, the spectrum in our setup is $m_H\ll m_A\simeq m_{H^\pm} = {\cal O}(100)\,\GeV$.
In Ref. \cite{Okawa:2020jea}, it is shown that this scalar spectrum can be achieved if the following three equations are satisfied:
\begin{align}
\lambda_4 + \lambda_5 &\approx - 2m_{H^\pm}^2/v^2 \,, \label{eq:cond1}\\
\lambda_4 - \lambda_5 &\approx 0 \,, \label{eq:cond2}\\
\lambda_3+\lambda_4+\lambda_5 &\approx 0 \,. \label{eq:cond3}
\end{align}
The first equation comes from Eq.\ (\ref{eq:MH}) by using $m_H \ll m_\Hch$ and holds up to ${\cal O}(m_H^2/v^2)$ corrections.
The second equation is necessary to respect the $T$ parameter constraint.
The third equation requires the consistency with the Higgs invisible decay bound, $\BR{h\to \text{inv}} \leq 0.13$\cite{ATLAS:2020cjb,Sirunyan:2018owy}, which limits the $H$-$H$-$h$ coupling $|\la_3+\la_4+\la_5|\lesssim 0.01$.
Eq. (\ref{eq:cond3}) should be fulfilled with an accuracy better than 1\%, since we have 
\begin{equation}
\la_3 \simeq - 2\la_4 \simeq -2\la_5 
    \simeq 2 m_\Hch^2/v^2 = {\cal O}(1) \,,
\end{equation}
from the three equations.
The values of $\lambda_{3,4,5}$ are thus fixed once a specific value of $m_\Hch$ is provided.

%%%%%%%%%%%%%%%%%%%%%%%%%%%%%%%
\subsubsection{Lepton portal couplings}
\label{sec:lepton-portal}
%%%%%%%%%%%%%%%%%%%%%%%%%%%%%%%
The new scalars also have Yukawa couplings to $\psi$ and SM lepton doublets $\ell_L$,
\begin{align}
\label{eq:ynu}
-{\cal L}_\ell= y^i_\nu \, \overline{\ell_{L}^i} \, \widetilde{\Phi_\nu} \psi_R + \hc \,,  
\end{align}
where $\widetilde{\Phi_\nu} \equiv i \sigma_2 \Phi_\nu^*$ and the Yukawa couplings $y^i_\nu$ $(i =e, \, \mu, \,\tau)$ are lepton flavor dependent complex parameters. 
For convenience, we choose the charged lepton mass basis, i.e.\ $\ell_L^i = \left[(U\nu_L)^i,\,e^i_L\right]^T$ where $e_L^i$ is a charged lepton in the mass basis and $U$ a neutrino mixing matrix. 
As discussed in Sec.~\ref{sec:DM}, the lepton portal couplings are responsible for the thermal DM production via $HH \to \nu \bar{\nu}$ annihilation.

In general, all three Yukawa couplings are non-vanishing. 
However, if more than one couplings have a comparably large magnitude, sizable charged lepton flavor violation is induced and the model will be ruled out.
In this paper, therefore, we focus on three particular flavor structures:
\begin{enumerate}
\renewcommand{\labelenumi}{(\roman{enumi})}
    \item $|y^e_\nu| \gg |y^{\mu,\tau}_\nu|$ (electrophilic case).
    \item $|y^\mu_\nu| \gg |y^{e,\tau}_\nu|$ (muonphilic case),
    \item $|y^\tau_\nu| \gg |y^{e,\mu}_\nu|$ (tauphilic case).
\end{enumerate}
These structures would be obtained by e.g.\ assigning a corresponding $U(1)$ lepton flavor charge to $\psi$, although in this paper we do not identify its UV origin and we just take those structures for the phenomenological purpose.
In the following, we assume $y^i_\nu$ to be real for simplicity. The relevant results would not change even if the Yukawa couplings are complex.

%%%%%%%%%%%%%%%%%%%%%%%%%%%%%%%
\subsubsection{Theoretical bounds}
\label{sec:th-bound}
%%%%%%%%%%%%%%%%%%%%%%%%%%%%%%%
As discussed above, the scalar quartic couplings are of ${\cal O}(1)$ to realize a light enough $H$.
These large couplings lead to Landau poles at low energy scales. 
In Ref.~\cite{Iguro:2022tmr} the authors study perturbativity bounds by analyzing the one-loop renormalization group evolution of the quartic couplings, and found that the charged scalar should be lighter than 350\,GeV to respect the perturbativity of the model up to 10\,TeV.
Note that other theoretical constraints, such as the bounded-from-below condition and absence of the charge-breaking vacuum, are satisfied in parameter space of our interest.

%%%%%%%%%%%%%%%%%%%%%%%%%%%%%%%%%%%%%%%%%%%%%%%%%
\subsection{Singlet extension}
\label{sec:extension}
%%%%%%%%%%%%%%%%%%%%%%%%%%%%%%%%%%%%%%%%%%%%%%%%%
One might wonder if the model presented in Sec.~\ref{sec:minimal} (referred to as the minimal model) can be ameliorated in a simple way for the theory to be valid at higher scales.
Indeed we can consider an extended model with a new $Z_2$-odd singlet scalar $S$ so that the fine-tuning between $\mathcal{O}(1)$ quartic couplings and the perturbativity constraint are significantly relaxed~\cite{Okawa:2020jea}.

Let us add the extra terms,
\begin{equation}
    \Delta {\cal L} = \frac{1}{2} (\partial_\mu S)^2 - \frac{1}{2} m_S^2 \, S^2 
    - [A_S \, \Phi^\dagger \Phi_\nu S + \hc] \,,
    \label{eq:L_singlet}
\end{equation}
to the minimal model.
After the EW symmetry breaking, $H$ and $S$ are mixed as
\begin{align}
    \label{eq;mixing}
    \begin{pmatrix}
        H \\ S
    \end{pmatrix}
    = 
    \begin{pmatrix}
        \cos\theta & -\sin\theta\\
        \sin\theta & \cos\theta
    \end{pmatrix}
    \begin{pmatrix}
        h_2 \\ s
    \end{pmatrix}\,,
\end{align}
where $\tan(2\theta)=2A_Sv/(m_H^2-m_S^2)$ and $h_2$ and $s$ are the mass eigenstates. 
The scalar masses are given by 
\begin{equation}
    m_{h_2}^2 = \frac{\cos^2\theta \, m_H^2 - \sin^2\theta \, m_S^2}{\cos(2\theta)}\,, \quad 
    m_s^2 = \frac{\cos^2\theta \, m_S^2 - \sin^2\theta \, m_H^2}{\cos(2\theta)} \,.
\end{equation}
One can see that $s$ can be arbitrarily light by setting $m_S \simeq \tan\theta \, m_H$, while $h_2$ (and the other scalars $H^\pm$ and $A$) can reside above the EW scale without any strong restriction on the quartic couplings $\la_{3,4,5}$.
In this extension, $s$ is DM, if lighter than $\psi$, and couples to $\psi$ through the scalar mixing $\theta$.

The $A_S$ term, which generates the scalar mixing, also induces a new decay mode $h\to ss$. 
This contribution to the Higgs invisible decay is constrained by experiments, and the current LHC bound corresponds to $\theta \leq {\cal O}(0.1)$, which will be improved to $\theta \lesssim 0.03$ in future, combining with direct searches for the charged scalar at the high luminosity (HL)-LHC~\cite{Iguro:2022tmr}. 
Note that, in principle, the Higgs invisible decay bound can be significantly relaxed by adding a new quartic interaction $S^2(\Phi^\dagger\Phi)$, which forms another contribution to the $h\to ss$ decay, at the cost of an additional 1\% level tuning between this new coupling and $A_S$. 
We do not pursue this possibility because only adding Eq. (\ref{eq:L_singlet}) can resolve all issues present in the minimal model.

In the singlet extension, 
the interaction of DM to $\psi$ and neutrinos is suppressed by a factor of $\theta$.
This means that the DM annihilation into neutrinos, which is responsible for the DM thermal production, is suppressed. 
The result concerning the DM production is easily translated from that of the minimal model by making $y_\nu^i \to y_\nu^i \sin\theta$ and $m_H \to m_s$ replacement.
Other DM constraints, such as those arising from the direct detection, are very weak in the singlet extension, mainly because all relevant processes are suppressed at least by a factor of $\theta^2$ at the amplitude level.
Therefore, in the rest of this paper, we only focus on the minimal model unless otherwise noticed.

%%%%%%%%%%%%%%%%%%%%%%%%%%%%%%%%%%%%%%%%%%%%%%%%%%
\section{Dark matter physics}
\label{sec:DM}
%%%%%%%%%%%%%%%%%%%%%%%%%%%%%%%%%%%%%%%%%%%%%%%%%%
$H$ and $\psi$ are good DM candidates. 
In Refs.~\cite{Okawa:2020jea, Iguro:2022tmr}, it is assumed that $\psi$ is lighter than $H$ and is DM. 
Here, we study the DM physics in the opposite case, i.e. $H$ is lighter than $\psi$ and is DM.

%%%%%%%%%%%%%%%%%%%%%%%%%%%%%%%%%%%%%%%%%%%%%%%%%%
\subsection{Relic density}
\label{sec:relic}
%%%%%%%%%%%%%%%%%%%%%%%%%%%%%%%%%%%%%%%%%%%%%%%%%%
We assume that DM is thermally produced, so that the DM abundance is determined by its annihilation cross section. 
Our DM $H$ annihilates into charged leptons, pions and photons in the sub-GeV region with its coupling to the SM Higgs boson, namely the Higgs portal coupling, 
\begin{equation}
    {\cal L} \supset - \frac{\la_{345}v}{2} H^2 h \,,
    \label{eq:higgs-portal}
\end{equation}
where we define 
\begin{equation}
    \la_{345} \equiv \la_3 + \la_4 + \la_5 \,.
    \label{eq:la345}
\end{equation}
It is known that the size of $\la_{345}$ required by the thermal DM production is too large to be consistent with the Higgs invisible decay bound.
Thus, we do not consider this negligible contribution from $\la_{345}$ in the DM production.

DM also annihilates into neutrinos with the lepton portal couplings $y_\nu^i$. 
The cross section is given by \cite{Toma:2013bka,Giacchino:2013bta,Kawamura:2020qxo}
\begin{equation}
(\sigma v_{\rel})_{HH\to\nu_i\bar{\nu}_i} 
    \simeq \frac{(y_\nu^i/\sqrt{2})^4 m_H^6}{60\pi(m_H^2+m_\psi^2)^4} v_{\rel}^4 \,,
\label{eq:HH-nunu}
\end{equation}
where we ignore neutrino masses and show the leading term in the expansion of the DM relative velocity $v_{\rel}$.
Since the annihilation is $d$-wave dominant, 
the coupling value required for the thermal production tends to be larger than the fermion DM case \cite{Okawa:2020jea, Iguro:2022tmr}.
In our analysis, the DM abundance is numerically calculated by employing {\tt micrOMEGAs\_5\_2\_13}~\cite{Belanger:2018ccd}, which implements calculations of all relevant coannihilation processes.

%%%%%%%%%%%%%%%%%%%%%%%%%%%%%%%%%%%%%%%%%%%%%%%%%%
\subsection{Direct detection}
\label{sec:DD}
%%%%%%%%%%%%%%%%%%%%%%%%%%%%%%%%%%%%%%%%%%%%%%%%%%
In this model, the lepton portal couplings do not generate measurable elastic DM-nucleon scattering, since the first non-vanishing contribution arises from two-loop diagrams.
Instead, the Higgs portal coupling $\la_{345}$ (and loop diagrams with the weak gauge couplings and scalar quartic couplings) can make the sizable contribution.

When the contribution from the lepton portal couplings is ignored, 
the DM-nucleon scattering in this model is induced the same way as in the inert doublet DM model \cite{Ma:2006km, Barbieri:2006dq, LopezHonorez:2006gr}. 
At the tree level, the Higgs portal process is the only relevant contribution to the scattering. 
It is shown in Ref.~\cite{Abe:2015rja} that there are also irreducible loop contributions to the scattering, which are non-vanishing even when $\la_{345}=0$ and can alter the tree-level prediction.
The loop corrections in the DM scattering are formally captured by an {\it effective} shift of the Higgs portal coupling, i.e.\ $\la_{345} \to \la_{345} + \delta \la$, 
where $\delta \la$ contains not only one-loop vertex correction to $\la_{345}$ but also loop contributions to $HH\bar{q}q$, $HHG_{\mu\nu}G^{\mu\nu}$ and quark twist-2 operators (see Eqs.~(\ref{eq:sigmaSI}) and (\ref{eq:dellamN}) below).

The loop corrections $\delta\la$ depend on the nucleon species ($N=p,n$), and following Ref.~\cite{Abe:2015rja}, we have\footnote{In Ref.~\cite{Abe:2015rja}, they only show the neutron coupling $\delta\la$ explicitly, which corresponds to $\delta\la_n$ in our notation, but it is straightforward to apply their results for the proton case.}
\begin{equation}
    \delta \la_N 
    = \delta\Gamma_h(0) + \delta\la_{345}
    + \frac{m_h^2}{f_N} 
        \sum_{q=u,d,s} \Gamma^q_{\rm Box} f_q^N 
    + \frac{2}{9} \frac{m_h^2}{f_N} \Gamma^G_{\rm Box} f_g^N 
    + \frac{3}{4} \frac{m_h^2}{f_N} 
        \sum_q \left(\Gamma^q_{\rm t2}+\Gamma^{\prime q}_{\rm t2}\right) \left(q^N(2)+\bar{q}^N(2)\right) \,,
    \label{eq:dellamN}
\end{equation}
where the values for $f_q^N$, $q^N(2)$ and $\bar{q}^N(2)$ are shown in Table \ref{tab:fN}, and $f_N$ and $f_g^N$ are defined as 
\begin{equation}
    f_N=\frac{2}{9}+\frac{7}{9}\sum_{q=u,d,s} f_q^N \,,\quad
    f_g^N = 1-\sum_{q=u,d,s} f_q^N \,.
\end{equation}
$\delta\Gamma_h(q_h^2)$ (with $q_h$ the Higgs four-momentum) and $\delta\la_{345}$ are the vertex correction and counter term to $\la_{345}$. 
We impose the renormalization condition as $\delta\la_{345} = -\delta\Gamma_h(m_h^2)$ so that the vertex correction does not change the tree-level prediction for the Higgs invisible decay: $\la_{345}$ in Eq.~(\ref{eq:sigmaSI}) is the renormalized value at $q_h^2=m_h^2$.
The last three terms in Eq.~(\ref{eq:dellamN}) are the loop contributions, apart from the vertex correction, and we refer readers to Ref.~\cite{Abe:2015rja} for the definition and complete expressions, while noting that $\delta \la_N$ depends on four free parameters $m_{H,A,\Hch}$ and $\la_2$, but is independent of $\la_{345}$.
We have numerically confirmed that the difference between $\delta\la_n$ and $\delta\la_p$ is smaller than 1\% in parameter space of our interest.
We thus focus only on the DM scattering to neutron in the following. 

%%%%%%%%%%%%%%%%%%%%%%%%%%%%%%%%%%%%%%%%%%%%%%%%%%%%%%%%%%%
\begin{table}[t]
\begin{center}
\begin{tabular}{cc}
    \begin{minipage}{0.3\hsize}
    \centering
    \begin{tabular}{|c|c||c|c|}
        \hline 
        $f_u^n$ & 0.0110 & $f_u^p$ & 0.0153 \\ \hline  
        $f_d^n$ & 0.0273 &  $f_d^p$ & 0.0191 \\ \hline
        $f_s^n$ & 0.0447 &  $f_s^p$ & 0.0447 \\
    \hline 
    \end{tabular}
    \end{minipage}
\hspace{1cm}
    \begin{minipage}{0.3\hsize}
    \centering
    \begin{tabular}{|c|c||c|c|}
        \hline 
        $u^n(2)$ & 0.11 & $\overline{u}^n(2)$ & 0.034 
        \\ \hline  
        $d^n(2)$ & 0.22 & $\overline{d}^n(2)$ & 0.036 \\ \hline  
        $s^n(2)$ & 0.026 & $\overline{s}^n(2)$ & 0.026\\ \hline  
        $c^n(2)$ & 0.019 & $\overline{c}^n(2)$ & 0.019 \\ \hline  
        $b^n(2)$ & 0.012 & $\overline{b}^n(2)$ & 0.012 \\ 
        \hline 
    \end{tabular}
    \end{minipage}
\end{tabular}
\end{center}
    \caption{
    Left: Nuclear matrix elements for light quarks. The values correspond to the {\tt micrOMEGAs} default.
    Right: The second moments of quark PDFs for neutron, which are numerically evaluated at $\mu=m_Z$ \cite{Abe:2015rja} using the CTEQ PDFs \cite{Pumplin:2002vw}. For proton, $u^p(2)=d^n(2)$ and $d^p(2)=u^n(2)$ and the second moments of the other quarks and anti-quarks are the same as those for neutron.}
    \label{tab:fN}
\end{table} 
%%%%%%%%%%%%%%%%%%%%%%%%%%%%%%%%%%%%%%%%%%%

\begin{figure}[t]
\centering
    \includegraphics[width=0.55\textwidth]{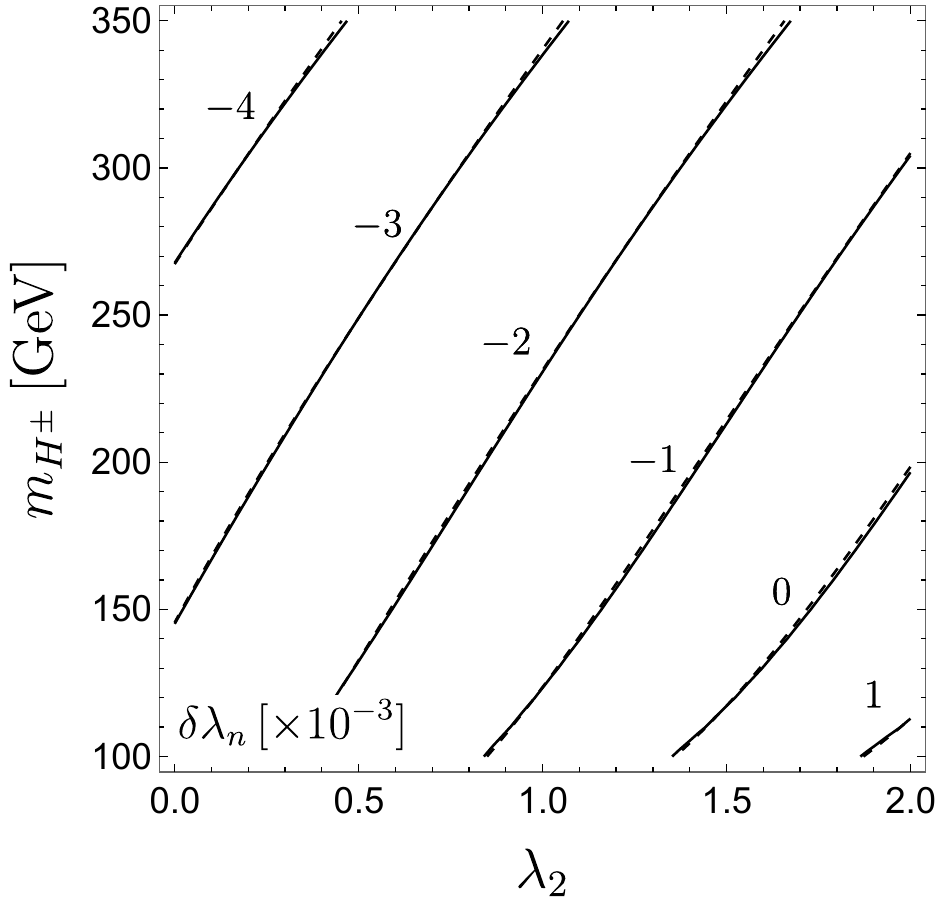}
    \caption{
    Contour lines for the loop corrections $\delta\lambda_n$ in unit of $10^{-3}$, which are evaluated using the full expression Eq.~(\ref{eq:dellamN}) (solid) and approximate formula Eq.~(\ref{eq:dellam_approx}) (dashed).
    }
    \label{fig:dellam}
\end{figure}

In Fig.~\ref{fig:dellam}, the black solid lines show the contours for the loop corrections $\delta \lambda_n$ with $m_\Hch=m_A$ and $m_H=10$\,GeV. 
One can see that $|\delta \lambda_n|$ is of ${\cal O}(10^{-3})$ in a large part of parameter space and amounts to 40\% of the possible maximal value of the tree-level coupling, i.e. $|\la_{345}|_{\rm max} \simeq 0.01$. 
The loop corrections can be dominant, depending on the value of $|\la_{345}|$ and the parameter region. 
It may be illuminating to provide a good approximate formula for $\delta \lambda_n$, which is given in the case of $m_\Hch=m_A$ by 
\begin{align}
\delta \lambda_n & \simeq 
    -0.00199 
    +(1.18\,m_H -6.3\,m_\Hch 
    -4.46\times10^{-3}\,m_\Hch^2)\times10^{-6}  
      \nonumber\\
    & \quad 
    + \la_2 \left( 
    0.00164
    + 2.57 /m_\Hch^2 
    + 5.76\times10^3/m_\Hch^4 
    %+ m_H^2 (-2.56983/m_\Hch^4 - 0.00163813 /m_\Hch^2) 
    \right) \,,
    \label{eq:dellam_approx}
\end{align}
where all masses are in the GeV unit.\footnote{Our formula is different from the one presented in Ref.~\cite{Abe:2015rja} mainly because we consider a different mass region. The formula in that paper predicts a non-vanishing value for $\delta\la_n$ when taking $m_H\to0$ and $\la_2\to0$ limit, but we figure out that $\delta\la_n$ is non-vanishing in that limit. We have tried several different fitting functions for $\delta\la_n$ and found that Eq.~(\ref{eq:dellam_approx}) better reproduces the numerical results in the parameter space of our interest.}
The $\la_2$ independent and dependent terms each agree with the numerical results with a precision better than 1$\%$, 
in the range of $0.1\,\GeV < m_H < 10\,\GeV$ and $100\,\GeV < m_\Hch < 350\,\GeV$ and $0< \la_2 < 2$.\footnote{$\la_2 >0$ is required by the bounded-from-below condition.}
The first line of the equation is always negative since $m_H \ll m_\Hch$, while the second line positive.
The black dashed lines in Fig.~\ref{fig:dellam} represent the contours from the approximate formula Eq.~(\ref{eq:dellam_approx}), which agree well with the numerical results evaluated using the full expression Eq.~(\ref{eq:dellamN}).

%==============================================
%==============================================

\begin{figure}[t]
\centering
    \includegraphics[width=0.46\textwidth]{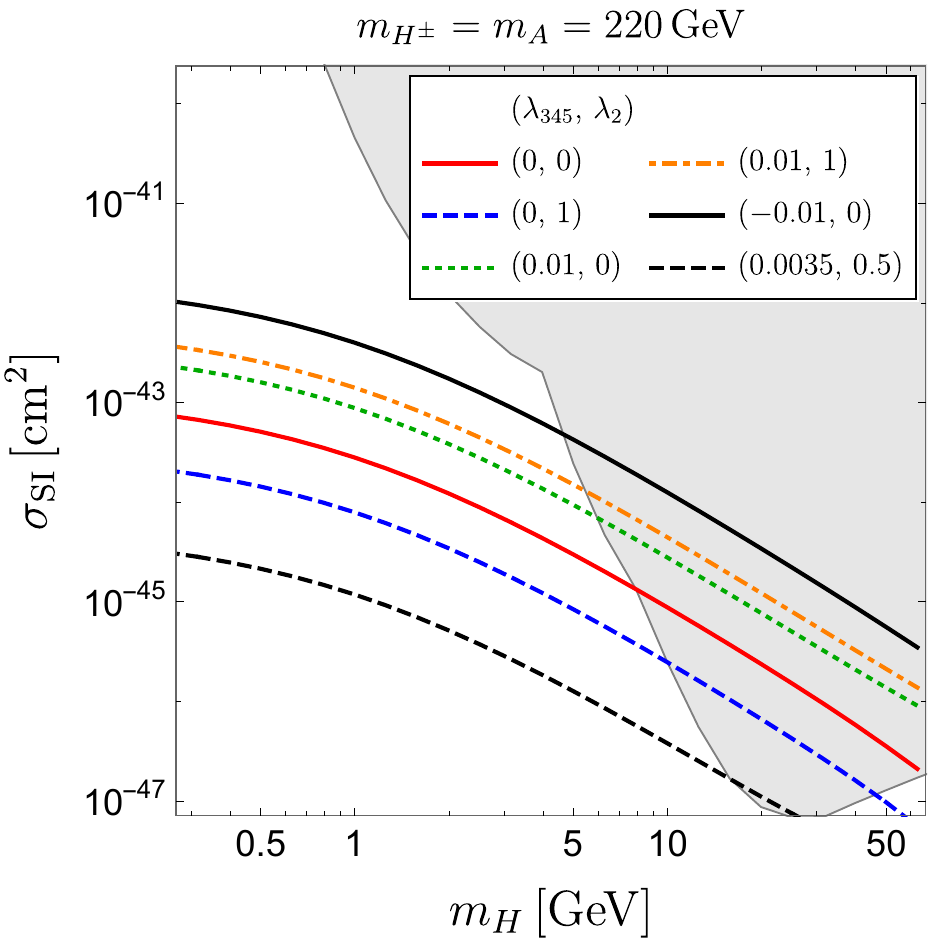}
    \includegraphics[width=0.48\textwidth]{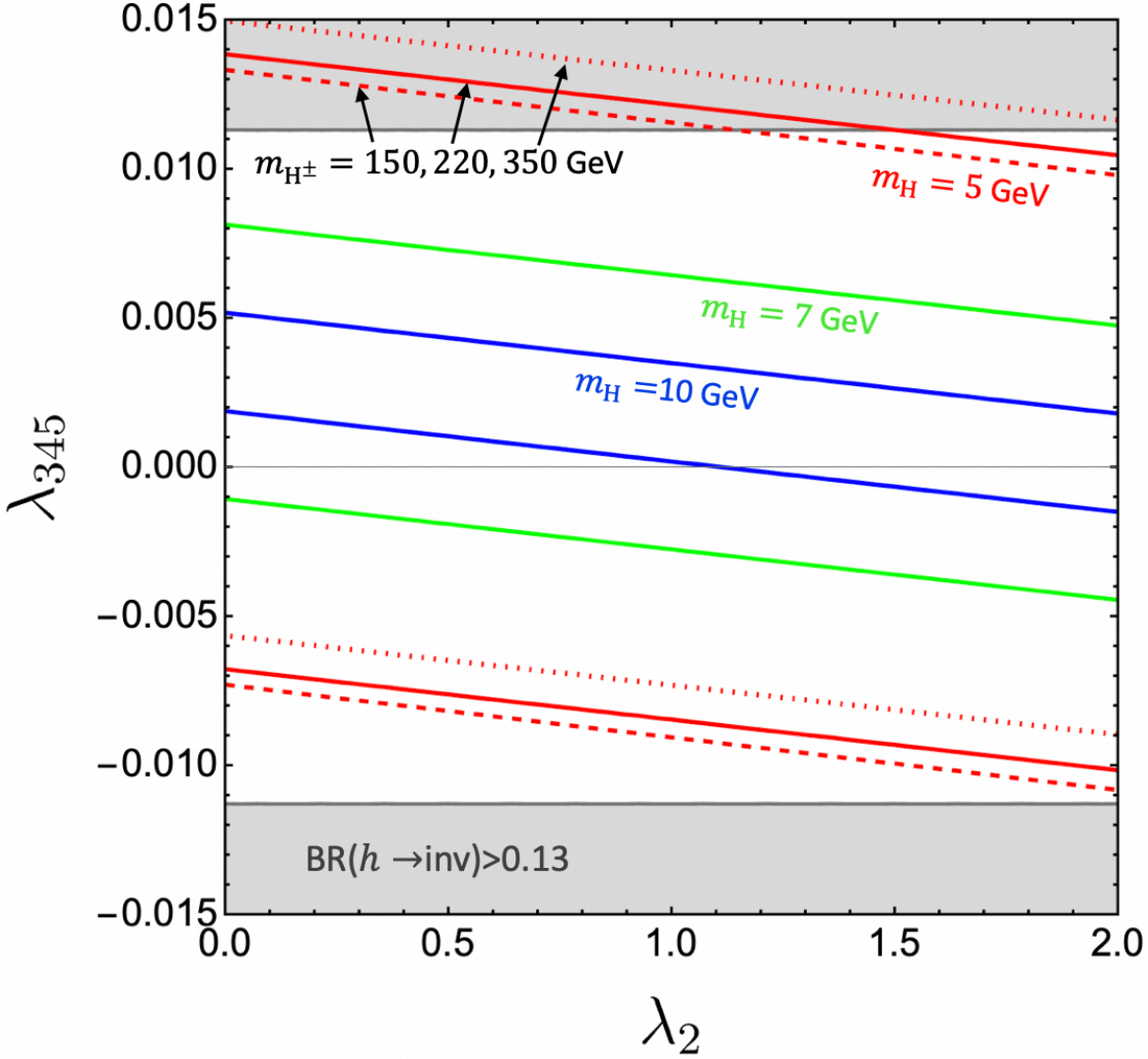}
    \caption{
    Left: Model predictions of $\sigma_{\rm SI}$ as a function of $m_H$ with $m_\Hch=m_A=220\,\GeV$ and several benchmark values of $\la_{345}$ and $\la_2$. 
    The shaded region is excluded by the current direct detection experiments.
    Right: The DM direct detection bounds in the parameter space, limited by the Higgs invisible decay (gray) and the bounded-from-below condition $\la_2 > 0$.
    The red lines correspond to $m_H=5\,\GeV$ with $m_{H^\pm}=150\,\GeV$ (dashed), 
    220\,GeV (solid) and
    350\,GeV (dotted), respectively.
    The green and blue solid lines are for $m_H=7\,\GeV$ and 10\,GeV with $m_{H^\pm}=220\,\GeV$.
    For each set of $m_H$ and $m_\Hch$ values, 
    the region between the two lines is allowed.
    }
    \label{fig:direct-detection}
\end{figure}

The spin-independent (SI) DM-neutron cross section is given by \cite{Abe:2015rja}
\begin{equation}
    \sigma_{\rm SI} = \frac{|\la_{345}+\delta\la_n|^2 m_n^4 f_n^2}{4\pi (m_H+m_n)^2 m_h^4} \,,
    \label{eq:sigmaSI}
\end{equation}
where $m_n$ is the neutron mass.
The values of $\la_{345}$ are free in our analysis as far as they satisfy the Higgs invisible decay bound.\footnote{$|\la_{345}|$ is fixed by the DM relic abundance in Ref.\cite{Abe:2015rja}.}
For reference, we show an estimate for $\sigma_{\rm SI}$, 
\begin{equation}
    \sigma_{\rm SI} 
        \simeq \frac{10^{-42}\,{\rm cm}^2}{(1+m_H/m_n)^2} \left(\frac{|\la_{345}+\delta\la_n|}{0.01} \right)^2 \,.
\end{equation}
It follows from this estimate that for $m_H \simeq 10\,\GeV$, $\sigma_{\rm SI}$ can be as large as $10^{-44}\,{\rm cm}^2$, which is large enough to be excluded by the current experimental results.
As $m_H$ gets lighter, $\sigma_{\rm SI}$ approaches an asymptotic value for a fixed $|\la_{345}+\delta\la_n|$. 
On the left panel in Fig.~\ref{fig:direct-detection}, 
we show the predictions for $\sigma_{\rm SI}$ with several benchmark values of $\la_{345}$ and $\la_2$ and with $m_\Hch=m_A=220\,\GeV$.
The black solid line corresponds to almost the maximum cross section.
The shaded region is excluded by the direct detection bounds, for which we used the current leading constraint in each mass region from DarkSide50 ($m_\psi \leq 5\,\GeV$) \cite{DarkSide-50:2023fcw}, PandaX-4T ($5\,\GeV \leq m_\psi \leq 9\,\GeV$) \cite{PandaX-4T:2021bab} and LZ ($9\,\GeV \leq m_\psi$) \cite{LZ:2022lsv}.

%==============================================
%==============================================

On the right panel in Fig.~\ref{fig:direct-detection}, we show the current direct detection constraints for $\la_{345}$ and $\lambda_2 $.  
The region of $|\la_{345}| \gtrsim 0.011$ is shaded in gray and excluded by the Higgs invisible decay at 95\% confidence level (CL).
The other lines (red, blue, green) correspond to the current sensitivity of the direct detection experiments for given sets of $m_H$ and $m_{H^\pm}$. 
The regions between two lines with the same color and style are allowed by the direct detection experiments.
The red lines represent the bounds for $m_H=5\,\GeV$ with $m_{H^\pm}=150\,\GeV$ (dashed), 220\,GeV (solid) and 350\,GeV (dotted), respectively. 
We see that the exclusion lines do not strongly depend on $m_{H^\pm}$.
The green and blue solid lines are for $m_H=7\,\GeV$ (green) and 10\,GeV (blue) with $m_{H^\pm}=220\,\GeV$ fixed. 
The allowed region for $m_H=10\,\GeV$ is much smaller than the one for $m_H=5\,\GeV$, mainly because the experimental limit on $\sigma_{\rm{SI}}$ weakens rapidly below 10\,GeV. 
A remarkable point is that $\la_{345}=0$, which is the easiest choice just to evade the Higgs invisible decay bound, is not always allowed by the direct detection results, taking the loop contributions into account. 
This suggests that the interplay between those two different searches provides an interesting window to test this model.
Before closing this subsection, we would like to repeat that the SI cross section discussed here is independent of the thermal DM production in our study.

%%%%%%%%%%%%%%%%%%%%%%%%%%%%%
\subsection{Indirect detection}
\label{sec:DM_IDD}
%%%%%%%%%%%%%%%%%%%%%%%%%%%%%
DM indirect detection experiments search for signals originated from DM annihilations into photon, charged particles, pions and neutrinos.
Since some experimental results limit the DM annihilation cross section to be far below the canonical thermal relic value in the sub-GeV region, 
it might be useful to look at the constraints in our case.

At the tree level, we have $HH\to\nu\bar{\nu}$ via the $t$-channel $\psi$ exchange. 
Given that the typical DM velocity in galaxies is so small as $v_\DM \sim 10^{-3}$, this process is strongly suppressed by the velocity (see Eq.\,(\ref{eq:HH-nunu})) and very weakly constrained.
The annihilation to the other particles occurs 
from the Higgs portal coupling and loop diagrams. 
Let us give a rough estimate for those cross sections below, assuming DM is lighter than 1 GeV. 
After all, we will see that the predicted DM annihilation cross section is too small for current and future planned experiments and indirect detection is irrelevant in our study.
This result clearly differs from that of the fermion DM scenario \cite{Okawa:2020jea,Iguro:2022tmr}.

Our DM annihilates into $e\bar{e}, \mu\bar{\mu}, \pi\pi, 2\gamma$ through the Higgs portal coupling. 
Given Eq.~(\ref{eq:higgs-portal}) and the low-energy Higgs couplings to the fermions and pions, 
\begin{equation}
    {\cal L} 
        = - \frac{h}{v} \left( \sum_{f=e,\mu} m_f \bar{f} f + \sum_{a=1,2,3} \left[\frac{2}{9} (\partial_\mu \pi^a)^2 - \frac{5}{6} m_\pi^2 (\pi^a)^2\right] \right) \,,
\end{equation}
we find the cross sections,
\begin{align}
(\sigma v)_{f\bar{f}}^{\text{Higgs}} 
    & \simeq \frac{1}{8\pi s} \frac{|\la_{345}|^2 m_f^2 (s-4m_f^2)}{m_h^4} \sqrt{1-\frac{4m_\pi^2}{s}} \,, \\
(\sigma v)_{\pi^a\pi^b}^{\text{Higgs}} 
    & \simeq \frac{\delta_{ab}}{16\pi s} \frac{|\la_{345}|^2}{m_h^4} \left(\frac{2}{9} s + \frac{11}{9} m_\pi^2 \right)^2 \sqrt{1-\frac{4m_\pi^2}{s}} \,, \\
(\sigma v)_{2\gamma}^{\text{Higgs}} 
    & \sim \frac{1}{16\pi s} \frac{|\la_{345}|^2 \la_{h\gamma}^2 s^2}{m_h^4} \,,
\end{align}
where $a,b=1,2,3$ and $\la_{h\gamma}$ denotes the off-shell Higgs to photon coupling evaluated at $p_h^2=s \ll m_h^2$ where $p_h$ is Higgs four-momentum.
Since $\la_{h\gamma} \sim \alpha/(4\pi) \ll 1$, the annihilation into pions is the biggest.
Taking the non-relativistic limit and massless pions (i.e. $s\simeq 4m_H^2 \gg 4m_\pi^2$) for reference, the cross section is estimated as 
\begin{equation}
(\sigma v)_{\pi\pi}^{\text{Higgs}} 
    \sim 2\times10^{-33}\,\text{cm}^3/\text{sec}
    \times 
    \left(\frac{|\la_{345}|}{10^{-2}}\right)^2 \left(\frac{m_H}{350\,\MeV}\right)^2 \,.
    \label{eq:HH-2pi}
\end{equation}
This is much smaller than the sensitivity of future gamma-ray telescopes -- for example, the projected 95\% CL upper limits at future e-ASTROGAM experiments are $10^{-29}\,\text{cm}^3/\text{sec}$ for $2\pi^0$ final state when DM mass is 350\,MeV (see Figure 4.5.2 in Ref.~\cite{e-ASTROGAM:2017pxr}).
We note that this equation is less valid when $m_H$ is close to $1$ GeV, because we apply the lowest order chiral perturbation calculation for the process at $\sqrt{s} \simeq 2m_H$ and higher order corrections could enhance or reduce the cross section. 
However, given that the lowest order estimate Eq.~(\ref{eq:HH-2pi}) is four orders of magnitude smaller than the future experimental reach, it is difficult to expect that our model is tested in this process even if the cross section calculation is refined.

%==============================
\begin{figure}
\centering
\includegraphics[viewport=50 570 560 760, clip=true, scale=0.65]{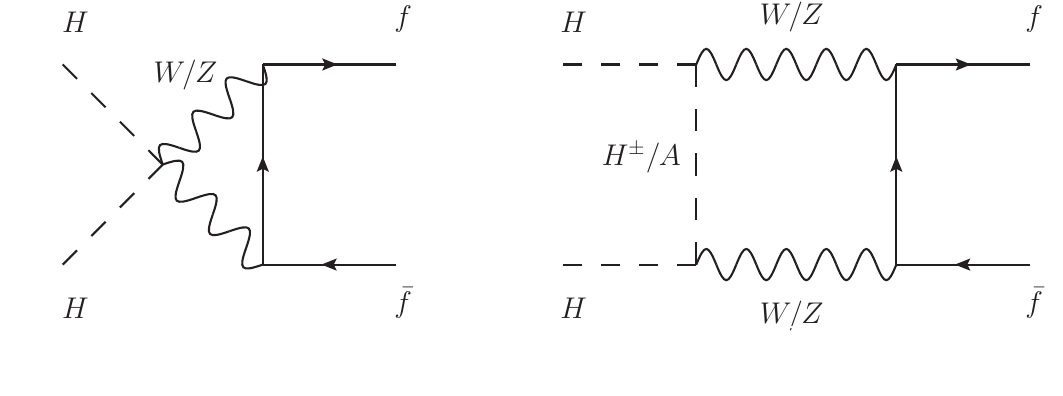}
\includegraphics[viewport=50 570 630 760, clip=true, scale=0.65]{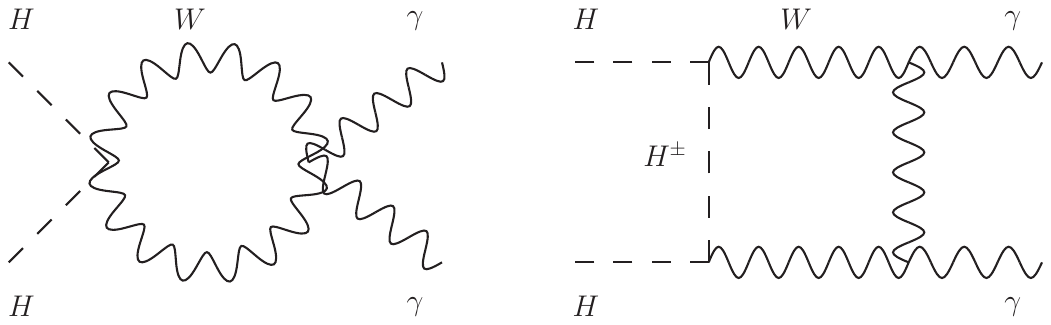}
\caption{Representative diagrams for one-loop DM annihilation.}
\label{fig:loop-annihilation}
\end{figure}
%==============================
Next, we discuss the processes induced by the loop corrections.
$H$ can annihilate into fermions and photon with $W$ and $H^\pm$ loop diagrams via the weak gauge interactions (see Fig.~\ref{fig:loop-annihilation}).
The cross sections are estimated as 
\begin{align} \label{eq314}
(\sigma v)_{f\bar{f}}^{\text{loop}} 
    & \sim \frac{1}{8\pi s} 
    \left(\frac{g^4 m_f}{16\pi^2 m_W^2}\right)^2 s 
       \sim 10^{-34}\,\text{cm}^3/\text{sec} 
       \times \left(\frac{m_f}{100\,\MeV}\right)^2,\\
(\sigma v)_{2\gamma}^{\text{loop}} 
    & \sim \frac{1}{16\pi s} 
    \left(\frac{g^2 e^2}{16\pi^2 m_W^2}\right)^2 s^2 
    \sim 10^{-33} \,\text{cm}^3/\text{sec} 
    \times \left(\frac{m_H}{\GeV}\right)^2 \,,
    \label{eq315}
\end{align}
where the $s$-wave contribution are only kept and evaluated at $s\simeq4m_H^2$. 
In Eq.~(\ref{eq314}), there is a helicity suppression factor, which is understood by the fact that the fermions in the final state must form the spin-0 state in the $s$-wave process. 
Clearly, both cross sections are too small to be within reach of the future gamma-ray telescopes.

%%%%%%%%%%%%%%%%%%%%%%%%%%%%%%%%%%%%%%%%%%%%%%%%%%
\subsection{$\Delta N_{\eff}$}
\label{sec:Neff}
%%%%%%%%%%%%%%%%%%%%%%%%%%%%%%%%%%%%%%%%%%%%%%%%%%
When $m_H \lesssim 20\,\MeV$, DM decouples from the thermal bath during or after the neutrino decoupling, which changes the neutrino-photon temperature ratio. 
This effect modifies the effective neutrino number $N_{\eff}$ that is constrained by the BBN and CMB measurements.

In Refs.~\cite{Boehm:2013jpa, Nollett:2014lwa, Heo:2015kra, Sabti:2019mhn, Sabti:2021reh}, the authors calculate a shift of $N_{\eff}$ from the SM prediction $\Delta N_{\eff}\equiv N_{\eff}-N_{\rm eff}^\SM$ as a function of dark sector particle mass, assuming that the dark sector particle coupled to electron or neutrino is in the thermal bath when becomes non-relativistic in the thermal history of the universe.
This result can be applied to the thermal relic DM case. 
For the real scalar DM coupled exclusively to neutrinos, the latest BBN+Planck bound reads $m_H \geq 4.3\text{--}5.6\,\MeV$ at $2\sigma$~\cite{Sabti:2021reh}, depending on the choice of the nuclear reaction rates for $d+d\to n+{^3{\rm He}}$ and $d+d\to p+{^3{\rm H}}$.
Since any set of the reaction rates in Refs. \cite{Pitrou:2020etk, Pisanti:2020efz, Yeh:2020mgl} reproduces the observed primordial deuterium abundance D$/$H$|_{\rm P}^{\rm obs}$ with good precision, we conservatively use the weakest limit $m_H \geq 4.3\,\MeV$ in this paper.
Note that this low mass bound also applies to the singlet extension.

%%%%%%%%%%%%%%%%%%%%%%%%%%%%%%%%%%%%%%%%%%%%%%%%%%
\subsection{Self-scattering}
\label{sec:SIDM}
%%%%%%%%%%%%%%%%%%%%%%%%%%%%%%%%%%%%%%%%%%%%%%%%%%
The quartic coupling $\la_2$ allows $H$ to scatter with itself at the tree level. 
The cross section is given, with a negligible contribution from the Higgs portal coupling, by
\begin{equation}
\sigma_{\rm self} = \frac{9\la_2^2}{32\pi m_H^2} \,.
\end{equation}
The presence of this tree-level self-scattering is in sharp contrast to the case of $\psi$ being DM, since $\psi$ has the self-scattering only at loop levels.

The sizable DM self-interaction affects cosmological structures on different scales, making a difference from the traditional collisionless DM paradigm and bringing constraints on the self-scattering cross section from cosmological observations (see e.g. Refs.~\cite{Tulin:2017ara, Adhikari:2022sbh} for reviews and references therein).
The observations have been made on various cosmological scales, and on the galaxy cluster scales, where the mean relative velocities of DM are 1000\,--\,2000\,km/s, the most severe constraints come from measurements of the core densities with strong gravitational lensing, providing $\sigma_\self/m_\DM < 0.13 \, {\rm cm}^2/{\rm g}$ in Ref.~\cite{Andrade:2020lqq} and $\sigma_\self/m_\DM < 0.35 \, {\rm cm}^2/{\rm g}$ in Ref.~\cite{Sagunski:2020spe}, both at 95\% CL.
The measurement on the galaxy group scales, with the average relative velocity of 1150\,km/s, gives a little weaker constraint $\sigma_\self/m_\DM < 1.1 \, {\rm cm}^2/{\rm g}$ \cite{Sagunski:2020spe}.
On the galaxy scales, where the DM relative velocity is of order of 10\,--\,100\,km/s, relatively large cross sections $\sigma_\self/m_\DM \simeq \text{1\,--\,10} \, {\rm cm}^2/{\rm g}$ are still allowed.
Given that the DM self-scattering is velocity independent in our model, we have an upper limit on $\la_2$,
\begin{equation}
    \la_2 \lesssim 0.10 \times \left(\frac{m_H}{10\,\MeV}\right)^{3/2} \left(\frac{(\sigma_\self/m_\DM)_{\rm exp}}{0.2\,\text{cm}^2/\text{g}}\right)^{1/2} \,,
\end{equation}
where $(\sigma_\self/m_\DM)_{\rm exp}$ is an experimental upper bound we choose.

%%%%%%%%%%%%%%%%%%%%%%%%%%%%%%%%%
\section{Other constraints}
\label{sec:others}
%%%%%%%%%%%%%%%%%%%%%%%%%%%%%%%%%
There are also experimental bounds on the heavy $Z_2$ odd particles $\psi$, $H^\pm$ and $A$, which directly and indirectly limit the DM parameter space when we assume the thermal production.
In this section, we study the constraints from the LHC and LEP experiments and SN 1987A.

%%%%%%%%%%%%%%%%%%%%%%%%%%%%%%%%%
\subsection{Constraints from the LHC experiments}
\label{sec:LHC}
%%%%%%%%%%%%%%%%%%%%%%%%%%%%%%%%%
The extra scalars are produced only in pairs through the EW gauge interactions in the proton collisions at the LHC. 
In Ref. \cite{Iguro:2022tmr}, the authors studied the searches for the extra scalar pair-productions at the LHC and showed that relevant bounds come only from $pp\to H^+H^-,\, AH$ productions followed by $H^\pm \to \psi \ell^\pm$ and $A \to Z H$ decays.
The bounds on the branching ratios of the $H^\pm$ and $A$ decays were derived by reinterpreting the results of mono-$Z$ \cite{ATLAS:2017nyv,CMS:2020ulv,ATLAS:2021gcn} and slepton searches \cite{CMS:2018eqb,CMS:2018yan,CMS:2019eln,ATLAS:2019lff,ATLAS:2019lng,CMS:2022syk} in our model.
The constraints derived in Ref.~\cite{Iguro:2022tmr} can be applied directly to this model, because the analysis in that paper was done without any assumption on the mass ordering of $H$ and $\psi$. 
Both particles were reconstructed as missing momenta, and this situation keeps holding in this paper.
We do not repeat the analysis here and simply use the results of Ref.~\cite{Iguro:2022tmr}.\footnote{See Sec.~3 of Ref. \cite{Iguro:2022tmr} for further details. Recently the ATLAS collaboration has reported a refined stau search~\cite{ATLAS:2023djh}.
This new search considerably improves the bounds and closes a gap in the light stau region.
However, we do not use this latest ATLAS bound due to the lack of available cross section data.}

\begin{figure}[t]
\centering
    \includegraphics[width=0.48\textwidth]{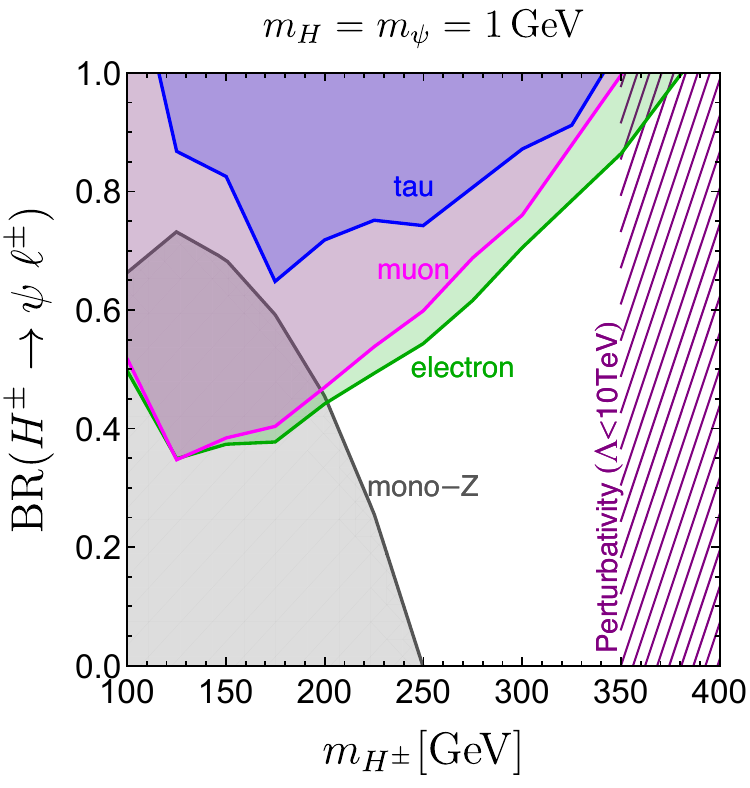}
    \includegraphics[width=0.48\textwidth]{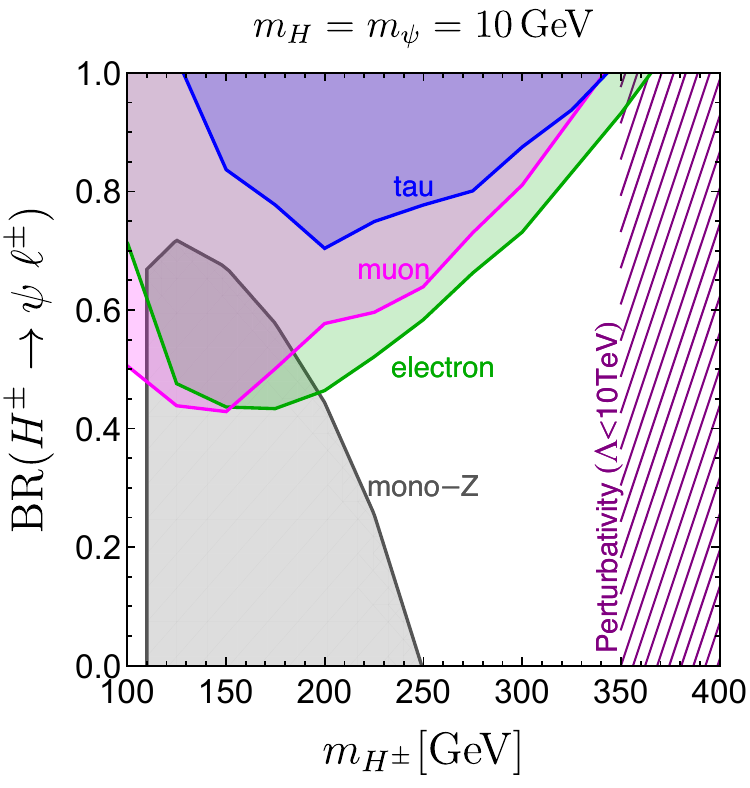}
    \caption{
    The LHC bounds from the slepton searches (green, magenta, blue) and mono-$Z$ search (gray) on the branching ratio of the charged scalar $\BR{H^\pm\to\psi\ell^\pm}$, with $m_\Hch=m_A$ and for $m_H=m_\psi=1\,\GeV$ (left) and $m_H=m_\psi=10\,\GeV$ (right). 
    The shaded regions are excluded.
    The mono-$Z$ bound applies to all flavor cases while  the slepton bounds depend on the flavor structure chosen.
    The hatched region predicts the low cutoff scales $\Lambda < 10\,$TeV from the perturbativity argument.
    }
\label{fig:LHC-limits}
\end{figure}

Fig.\,\ref{fig:LHC-limits} shows the constraints on the branching ratio of the charged scalar BR($H^\pm\to \psi \ell^\pm$), for $m_H=m_\psi=1$\,GeV (left) and 10\,GeV (right) and with $m_\Hch=m_A$, which are derived from the LHC Run 2 data with 139\,fb$^{-1}$ luminosity.
The shaded regions are excluded by the slepton and mono-$Z$ searches. 
While the bound from slepton search depends on the lepton flavor, the mono-$Z$ bound is common to all flavor structures. 
The hatched region predicts the low cutoff scale: $\Lambda <10\,$TeV.
When the cutoff scale is required to be above 10 TeV, the charged scalar mass has to be below $ 350\,\GeV$. 
The mono-$Z$ search indeed gives an {\it upper} bound on $\BR{A \to HZ}$, which is translated into the {\it lower} bound on $\BR{H^\pm\to \psi \ell^\pm}$ in Fig.~\ref{fig:LHC-limits}.  
The mono-$Z$ search excludes the region where $m_\Hch \le 250\,$GeV, when $\BR{H^\pm\to \psi \ell^\pm} \approx 0$, i.e. $y^i_\nu \approx 0$.
One may notice that there is a narrow allowed region in $100\,\GeV<m_\Hch<110\,\GeV$ when $m_H=10\,\GeV$.
This region is not always consistent with the thermal production in this model. 
As we will see later in Sec.~\ref{sec:result}, when $m_H=m_\psi=10\,\GeV$, the correct thermal production requires $y_\nu^i \simeq 0.15$, which corresponds to $\BR{H^\pm \to \psi \ell^\pm} \simeq 0.56$ for $m_{H^+}=110\,\GeV$. 
This branching ratio is compatible with the slepton searches in the electro and tauphilic cases, but is excluded in the muonphilic case. 
This conclusion might be changed by lifting $m_A=m_{H^\pm}$ and/or taking into account off-shell $W$-mediated three-body decays $H^\pm \to H W^* \to H f\bar{f}'$.\footnote{We have confirmed that this process only mildly reduces $\BR{H^\pm \to H \ell^\pm}$ for the $m_{H^\pm}=m_A$ case.}
The slepton searches provide upper limits on BR($H^\pm\to \psi \ell^\pm$).
Combining the slepton and mono-$Z$ bounds, 
$m_\Hch \lesssim 200\,\GeV$ is excluded in the electro and muon-philic scenarios, while a relatively large parameter space is still allowed in the tauphilic case.

%==============================================
%==============================================

%%%%%%%%%%%%%%%%%%%%%%%%%%%%%%%%%
\subsection{Constraints from the LEP experiments}
\label{sec:LEP}
%%%%%%%%%%%%%%%%%%%%%%%%%%%%%%%%%
In the electrophilic case, there is a Yukawa coupling, $ y^e_\nu \bar{\psi}_R H^+ e_L$. This coupling leads the mono-photon plus missing energy process, $e^+e^- \to \psi\bar{\psi} \gamma$, through the charged scalar exchange in the LEP experiments. 
In Ref.~\cite{Fox:2011fx}, this mono-photon search is studied in an effective field theory consisting of four-fermi electron-DM operators (as well as in simplified single mediator models), and gives upper limits on the Wilson coefficients.
In Ref.~\cite{Iguro:2022tmr}, integrating out the charged scalar and matching with their effective field theory, we translate the limit in Ref.~\cite{Fox:2011fx} into an upper limit on the electrophilic lepton portal coupling in our model:
\begin{equation}
|y_\nu^e| \leq \frac{m_\Hch}{240\,\GeV} \,.
\end{equation}
Although we considered $m_\psi < m_H$ in that paper, this bound can be applied to the present mass spectrum $m_H < m_\psi$, since $\psi$ decays only to $H$ and $\nu$ and its decay does not affect the observed single photon spectrum.

%%%%%%%%%%%%%%%%%%%%%%%%%%%%%%%%%
\subsection{Constraints from SN 1987A}
\label{sec:SN}
%%%%%%%%%%%%%%%%%%%%%%%%%%%%%%%%%
Light dark sector particles can be copiously produced in the core of supernovae (SNe) through the interactions with the stellar medium and, when escaping from the stars without absorption, carry the energy away. 
This extra energy release modifies the cooling rate during the burst and brings constraints on the interactions between the SM and dark sectors from the observation of SN 1987A.

In this model, the mediator $\psi$ is produced via $e^+ e^-$ and $\mu^+\mu^-$ annihilation \cite{Dreiner:2003wh, Dreiner:2013mua, Kadota:2014mea, Guha:2018mli, Manzari:2023gkt}. 
In the electrophilic (muonphilic) case, the lepton portal coupling $y_\nu^e$ ($y_\nu^\mu$) induces $e^+e^-\to\psi\bar{\psi}$ ($\mu^+\mu^-\to\psi\bar{\psi}$) with the tree level $H^\pm$ exchange, while no such a process is induced at the tree-level in the tauphilic case. 
In this paper, we cast SN bounds on Wilson coefficients of electron-$\psi$ and muon-$\psi$ four-fermi operators, derived in Ref.~\cite{Manzari:2023gkt}, onto 
constraints on $y_\nu^e$ and $y_\nu^\mu$. 
More precisely we take as a benchmark constraint the solid lines in Fig.\,3 of that paper, 
which are obtained using numerical input from SFHo-18.8 SN simulation \cite{Bollig:2020xdr}.
It should noted that the full DM mass dependence of the SN bounds is provided only for the $V\otimes V$ operators in Ref.~\cite{Manzari:2023gkt}, while the $(V-A)\otimes(V+A)$ operator arises in our model after Fierz transformation. 
We see, however, in Table 1 of that paper that this operator difference affects the SN bound by 50\% at most.
This effect is smaller than the one stemming from the uncertainty  as to the SN simulation, which changes the cooling bound by a factor of 2 (see Fig.~3 of that paper).

A remarkable difference from the literature is that in our model, $\psi$ produced in the core quickly decays into $H$ and $\nu$.
The neutrino produced from the decay easily loses the kinematic energy by scattering to the dense SN medium, and gets reabsorbed.
On the other hand, $H$ hardly scatters off the SN medium and thus carries the energy away from the SN.
We simply assume that half of the $\psi$ energy is released from the SN.\footnote{This assumption would not be justified when $\psi$ is degenerate in $H$.
We found, however, that in the parameter space where the SN bound is relevant, $m_H \ll m_\psi$ is satisfied.}
This half energy release is taken into account by rescaling the upper boundary of the $\Lambda^\eff_{e,\mu}$ bound in Ref.~\cite{Manzari:2023gkt} by a factor of $2^{1/4}$ for a fixed $m_\psi$.
The lower boundary of the $\Lambda^\eff_{e,\mu}$ bound, which usually appears as a result of absorption of the produced $\psi$ while passing through the SN, vanishes in the present study.
This is because $\psi$ decays away before it scatters, and the decay product $H$ little scatters off the SN medium.

As another remark, we add that DM production from neutrino annihilation also brings the SN 1987A bound into play.
Recently Ref.~\cite{Fiorillo:2022cdq} 
derives that bound on a Majoron which is resonantly produced from annihilation of neutrinos abundant in the SN.
Ref.~\cite{Manzari:2023gkt} also discusses the SN bound from the neutrino-initiated DM production with an s-channel vector mediator $Z^\prime$ and in an effective four-fermi theory, corresponding to the scenario with a mediator heavy enough to be integrated out in the relevant cooling process. 
In our model, $\psi$ ($H$) would be produced through $\nu\overline\nu \to \psi \overline \psi\,(HH)$ via t-channel $H$ ($\psi$) exchange, thereby increasing the DM emission rate.
Although the increase of the emission rate would strengthen the SN bound, any existing analysis cannot apply to our scenario, because of completely different production and decay channels for the DM and mediator. 
We do not analyze that process, which is beyond the aim of this paper.
The detailed study to implement the neutrino-involved process will be done in future.

%%%%%%%%%%%%%%%%%%%%%%%%%%%%%%%%%%%%%%%%%%%%%%%%%
\section{Results}
\label{sec:result}
%%%%%%%%%%%%%%%%%%%%%%%%%%%%%%%%%%%%%%%%%%%%%%%%%
In this section, we discuss the model predictions, based on the constraints studied in Sec.~\ref{sec:DM} and Sec.~\ref{sec:others}. 
We also compare our results with those presented in the previous works \cite{Okawa:2020jea,Iguro:2022tmr} where $\psi$ is DM, i.e.~$m_\psi<m_H$. 
Hereafter, we refer to the case of $H$ being DM as the scalar DM case, and the case of $\psi$ being DM as the fermion DM case for clarity.

\begin{figure}[p]
\centering
\includegraphics[width=0.49\textwidth]{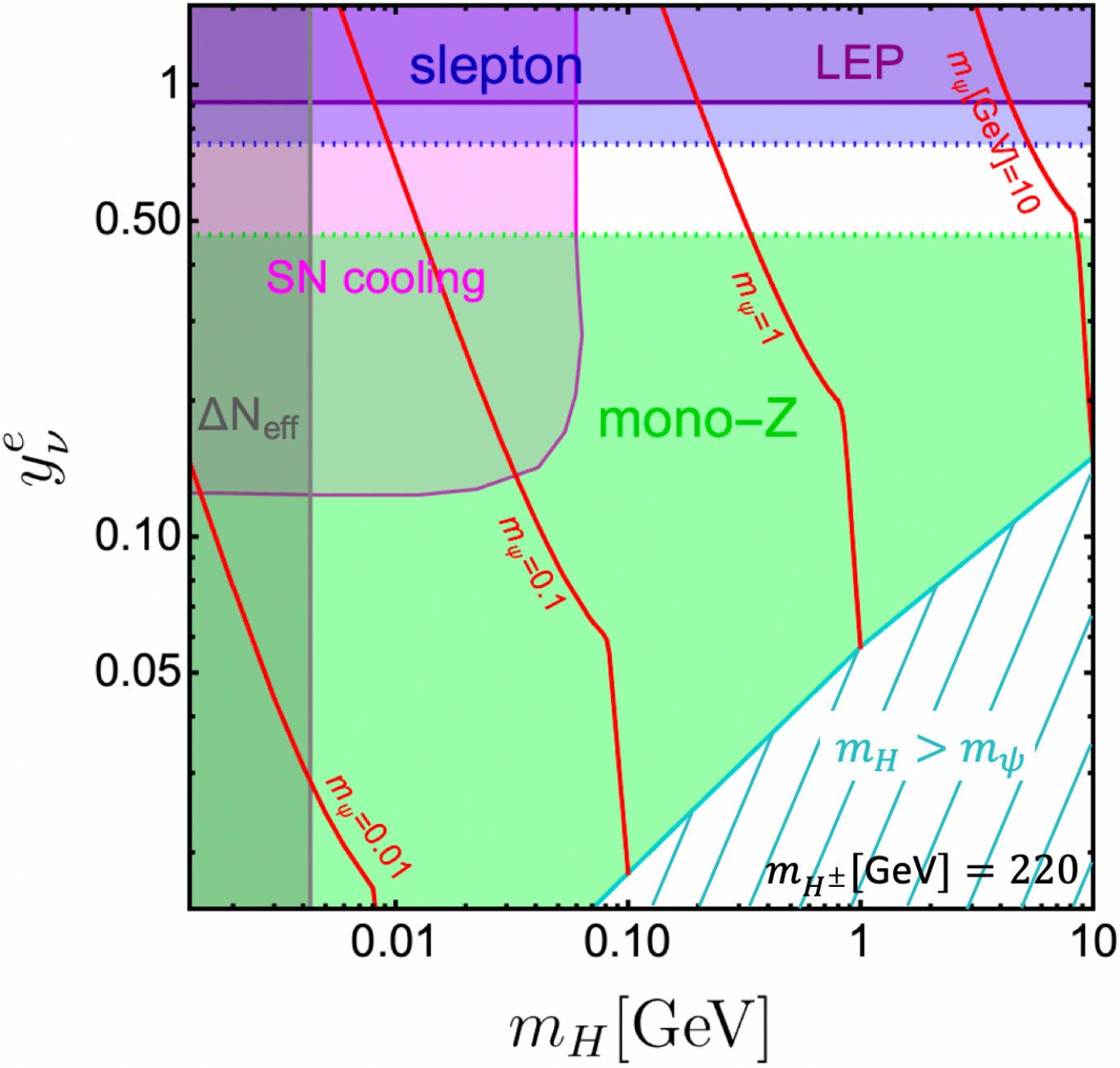}
\includegraphics[width=0.49\textwidth]{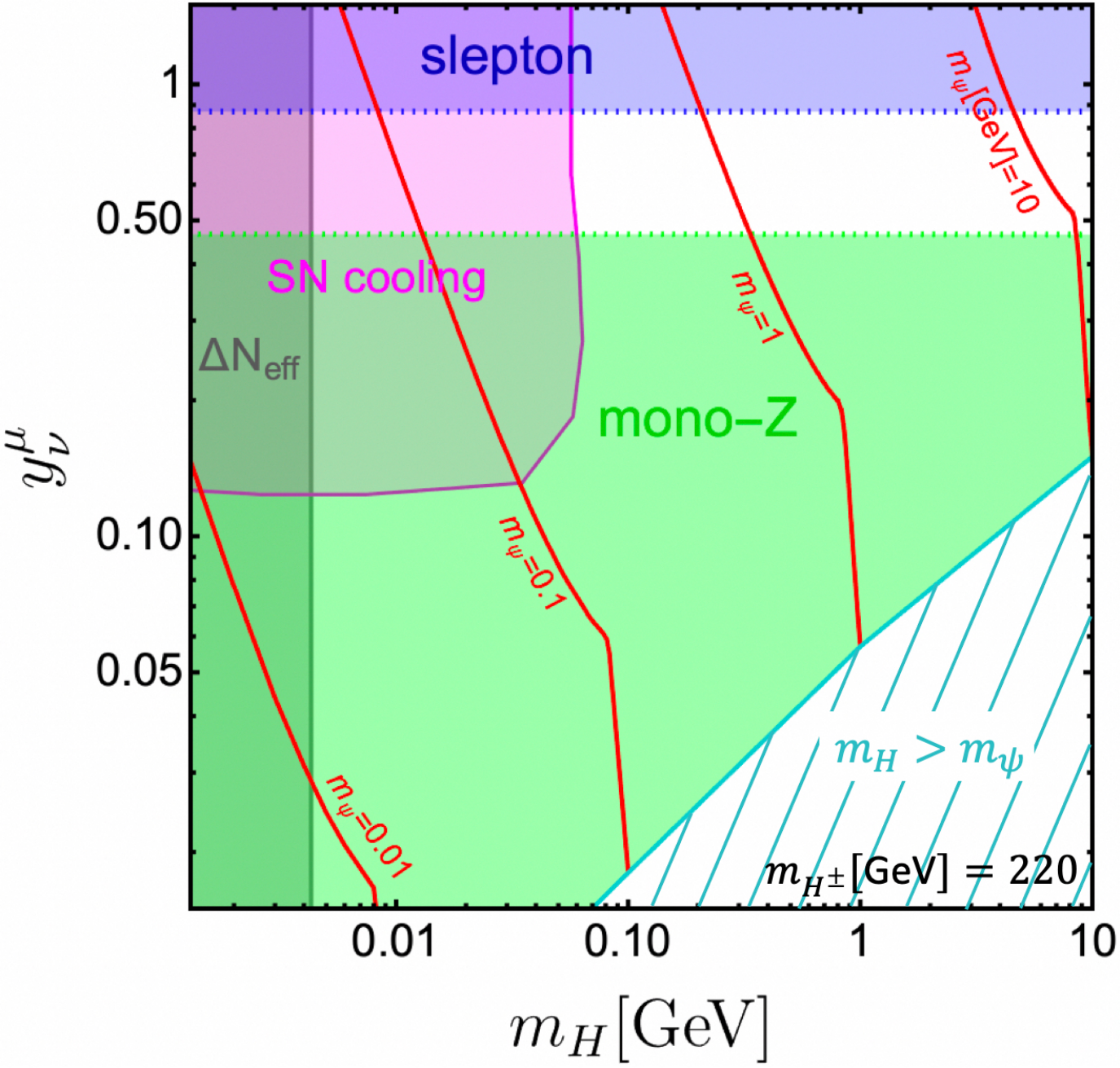}\\[0.3cm]
\includegraphics[width=0.49\textwidth]{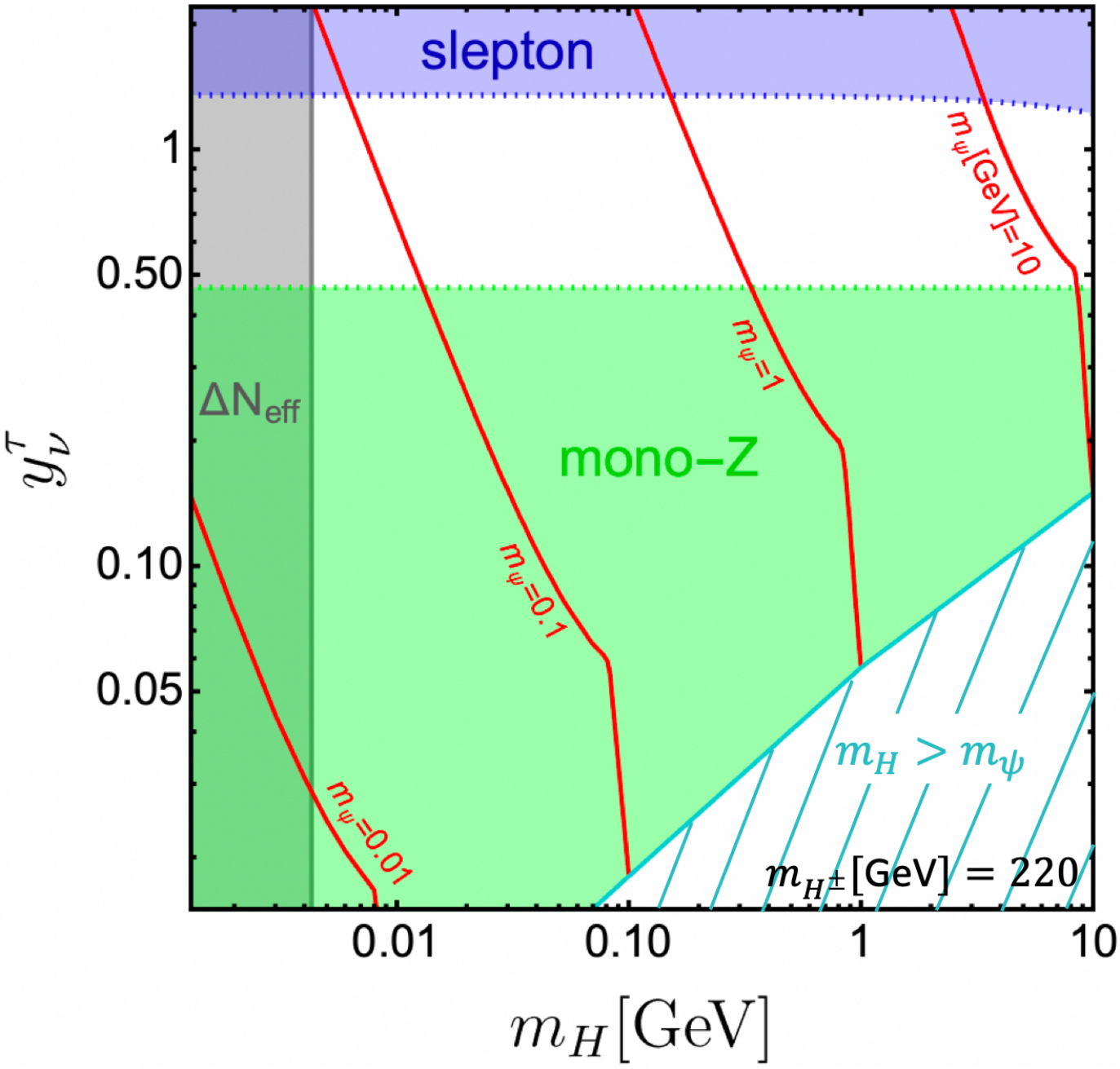}
    \caption{
    The experimental constraints from the $\Delta N_\eff$, LHC, LEP, SN1987A with $m_{H^\pm}=220\,$GeV 
    for the electrophilic (upper left), muonphilic (upper right) and tauphilic (lower) cases.
    The shaded regions are excluded.
    The red lines show the contours for the values of $m_\psi$ that can thermally produce the observed DM abundance. 
    In the cyan hatched region, $H$ cannot be thermal relic DM, 
    because any $m_\psi$ value does not reproduce the observed abundance in the thermal relic scenario, unless $m_H > m_\psi$, namely $\psi$ is DM, or non-thermal production is considered.
    }
    \label{fig:result}
\end{figure}

In our model, the DM abundance is determined by three parameters, $m_H$, $m_\psi$, and $y_\nu$, and independent of $m_{H^\pm}$ and the flavor structures. 
Figure \ref{fig:result} illustrates how the observed DM abundance can be thermally realized in the $(m_H, y_\nu^i)$ plane, with the red lines representing values of $m_\psi = 0.01, 0.1, 1$ and 10\,GeV.
Clearly, these red lines exhibit a different behavior from those in the fermion DM case \cite{Iguro:2022tmr} (see Fig.~6 in that paper). 
This is understood from the scaling of the DM annihilation cross section for $m_\DM \ll m_{\rm med}$, where $m_\DM$ and $m_{\rm med}$ denote masses of the DM and mediator in each case.
In this limit, $(\sigma v_\rel) \propto y_\nu^4 m_\DM^6/m_{\rm med}^8$ in the scalar DM case, while $(\sigma v_\rel) \propto y_\nu^4 m_\DM^2/m_{\rm med}^4$ in the fermion DM case. 
It follows from this scaling that with $m_{\rm med}$ fixed, $y_\nu \propto m_\DM^{-3/2}$ in the scalar DM case while $y_\nu \propto m_\DM^{-1/2}$ in the fermion DM case, since the thermal freeze-out production requires a constant cross section value, regardless of DM spin and mass.
Further in the scalar DM case, as $m_H \to m_\psi$, a coannihilation process $\psi \bar{\psi} \to \nu \bar{\nu}$ efficiently produces the DM abundance, because this process is not velocity suppressed. 
The efficient coannihilation contribution allows DM to be produced with a smaller $y_\nu^i$, and this is why  in Fig.\,\ref{fig:result} the slope in the mass degenerate region ($m_H \simeq m_\psi$) is steeper than the mass split region ($m_H \ll m_\psi$). 
In the cyan hatched region, $H$ cannot be thermal relic DM, because any $m_\psi$ value does not reproduce the observed abundance in the thermal relic scenario, unless $m_H > m_\psi$ and $\psi$ is DM.

%===================================================
%===================================================

We also show in Fig.\,\ref{fig:result} the constraints from the slepton (blue) and mono-$Z$ (green) searches, $\Delta N_{\rm eff}$ (gray), LEP (purple) and SN 1987A (magenta) for each flavor structure.
The shaded regions are excluded. 
Of these constraints, the $\Delta N_\eff$ bound depends only on $m_H$ and restricts it to be $m_H \geq 4.3\,\MeV$, which forms another difference from the fermion DM case in which the $\Delta N_\eff$ bound is $m_H \geq 10\,\MeV$.
The other constraints are all described by four parameters, $m_H$, $m_\psi$, $y_\nu$ and $m_\Hch (=m_A)$, except the DM direct detection, which we do not consider in this section, since it depends on $\lambda_{345}$ and $\lambda_{2}$ and can always be evaded by appropriately choosing those two parameter values (see Fig.~\ref{fig:direct-detection}).
Among the four parameters, we fix the charged scalar mass at $m_\Hch=m_A=220\,\GeV$ in Fig.~\ref{fig:result}, thereby making it easy to compare with the fermion DM results.

The LHC constraints exclude a large part of parameter space for this charged scalar mass $m_\Hch=220\,\GeV$. 
These constraints become significantly weaker (stronger) as $H^\pm$ gets heavier (lighter).
Concretely, we see in Fig.\,\ref{fig:LHC-limits} that in the electro and muonphilic cases and for $m_{H^\pm} \lesssim 200\,\GeV$, the combination of the slepton and mono-$Z$ searches rule out all possible $\BR{H^\pm \to \psi \ell^\pm}$ value, which in turn rules out any $y_\nu^i$ value. 
Whereas, when $m_{H^\pm}\ge250\,$GeV is chosen, the mono-$Z$ search does not work and even $\BR{H^\pm \to \psi \ell^\pm} = 0$ is allowed, and as a result the mono-$Z$ exclusion region in Fig.\,\ref{fig:result} completely disappears from the plots.
In the tauphilic case, the constraint is relatively weak and charged scalar can be a bit lighter, but only a limited parameter space around $y_\nu^\tau \simeq 1$ is viable when $m_\Hch \lesssim 200\,\GeV$.
All parameter spaces that are currently unexplored will be expected to be within reach of the future HL-LHC experiment \cite{Iguro:2022tmr}.

%%%%%%%%%%%%%%%%%%%%%%%%%%%%%%%%%%%%%%%%%%%%%%%%%%%%%%%%
\section{Summary}
\label{Sec:summary}
%%%%%%%%%%%%%%%%%%%%%%%%%%%%%
In this paper, we studied a light scalar DM model that features an inert scalar doublet $\Phi_\nu$ and singlet Dirac fermion $\psi$, which are both odd under a new global $Z_2$ symmetry while all SM fields are even. 
The DM originates in the inert doublet, which couples to the SM sector with quartic couplings to the SM Higgs doublet as well as Yukawa couplings to $\psi$ and the left-handed leptons which are referred to as the lepton portal couplings. 
We consider the case that the DM mass is below 10\,GeV, which is realized by adjusting three quartic couplings as shown in Eqs.~(\ref{eq:cond1})-(\ref{eq:cond3}). 
In this mass region, the DM can be thermally produced by its pair annihilation into neutrinos, which is mediated by $\psi$.

Focusing on the parameter space where the lightest inert scalar $H$ is DM, we discussed the DM physics and experimental constraints on the extra scalars and compared our results with those studied in the case of $\psi$ being DM \cite{Okawa:2020jea, Iguro:2022tmr}. 
The differences are summarized as follows:
\begin{itemize}
    \item In the thermal production (Sec.~\ref{sec:relic}), the velocity suppression and different mass scaling in the DM annihilation result in a larger lepton portal coupling than in the fermion DM case.

    \item In the direct detection (Sec.~\ref{sec:DD}), 
    the lepton portal couplings do not generate any relevant contribution to the elastic DM-nucleon scattering. This makes a striking contrast to the fermion DM case, where the lepton portal couplings are responsible for both the DM production and direct detection, and thus correlate them closely. 
    In the scalar DM case, the weak gauge and quartic interactions instead make the leading contribution from the loop diagrams, which depends on the Higgs portal coupling and DM self-coupling but decouples from the DM production.

    \item The indirect detection (Sec.~\ref{sec:DM_IDD}) does not provide useful probes of the scalar DM, while the monochromatic neutrino flux from the galactic DM annihilation is a good probe in the fermion DM case.

    \item The $\Delta N_\eff$ bound (Sec.~\ref{sec:Neff}) restricts the DM mass to be heavier than 4.3\,MeV, in contrast to 10\,MeV for the fermion DM case.

    \item The scalar DM has the self-scattering at the tree level (Sec.~\ref{sec:SIDM}), while it is loop induced in case $\psi$ is DM.
\end{itemize}
On the other hand, the collider bounds are basically the same as in the fermion DM case, given that both $H$ and $\psi$ are supposed to be missing.

In closing, it is worth mentioning that adding the inert doublet with the light $H$ triggers strong first order EW phase transition \cite{Chowdhury:2011ga, Borah:2012pu, Gil:2012ya, Cline:2013bln, Blinov:2015sna, Blinov:2015vma}, pointing to the EW baryogenesis as a promising baryogenesis scenario. 
It is unfortunate that the EW baryogenesis mechanism does not operate with the model considered in this paper, due to the absence of new CP violating couplings to the Higgs field.
This encourages us to extend the model to incorporate other renormalizable CP-violating interactions to enable the EW baryogenesis as discussed in Refs. \cite{Cline:2017qpe, Fernandez-Martinez:2020szk, Fernandez-Martinez:2022stj}.

%%%%%%%%%%%%%%%%%%%%%%%%%%%%%%%%%%%%%%%%%%%%%%%%%%%%%%%%
\section*{Acknowledgements}
%%%%%%%%%%%%%%%%%%%%%%%%%%%%%%%%%%%%%%%%%%%%%%%%%%%%%%%%
%---------------------------------------------------------------------------
S.I. appreciates Robert Ziegler for the fruitful discussion on the SN bound.
S.O.\ is grateful to Tomohiro Abe for clarification of the loop effects on the DM-nucleon scattering in the inert doublet model.
S.I. would like to thank support from a Mar\'ia de Maeztu grant for a visit to 
the Institute of Cosmos Sciences, Barcelona University,
where he stayed during the initial stage of this project.
S.I. enjoys the support from the Deutsche Forschungsgemeinschaft (DFG, German Research Foundation) under grant 396021762-TRR\,257.
%%%
S.O. acknowledges support from a Maria Zambrano fellowship, from the State Agency for Research of the Spanish Ministry of Science and Innovation through the ``Unit of Excellence Mar\'ia de Maeztu 2020-2023'' award to the Institute of Cosmos Sciences (CEX2019-000918-M) and from PID2019-105614GB-C21 and 2017-SGR-929 grants.
%---------------------------------------------------------------------------
The work of Y.O. is supported by Grant-in-Aid for Scientific research from the MEXT, Japan, No. 19K03867.
%%%
%---------------------------------------------------------------------------

%%%%%%%%%%%%%%%%%%%%%%%%%%%%%%%%%%%%%
%%%%%%%%%%%%%%%%%%%%%%%%%%%%%%%%%%%%%
\bibliographystyle{utphys28mod}
\bibliography{ref,ref_lightDM}
\end{document}